\documentclass[aps,prl,groupedaddress,amsmath,twocolumn,floatfix,showpacs]{revtex4}
\usepackage{graphicx}
\usepackage{bm}
\usepackage{times}
\usepackage{txfonts}

\renewcommand{\Re}{\text{\,Re\,}}
\renewcommand{\Im}{\text{\,Im\,}}

\newcommand{\so}{s_0}
\newcommand{\So}{S_0}

\newcommand{\Hetrimer}{$^4$He$_3$}
\newcommand{\Hedimer}{$^4$He$_2$}
\newcommand{\Heatom}{$^4$He}

\newcommand{\expect}[1]{\left\langle #1 \right\rangle}

\begin{document}

\title{Matter Wave Diffraction from an Inclined Transmission Grating:
  Searching for the Elusive $^4$He Trimer Efimov State}

\author{R.~Br{\"u}hl}
\author{A.~Kalinin}
\author{O.~Kornilov}
\author{J.~P.~Toennies}
\affiliation{Max-Planck-Institut f{\"u}r Str{\"o}mungsforschung,
  Bunsenstra{\ss}e 10, 37073 G{\"o}ttingen, Germany}
\author{G.~C.~Hegerfeldt}
\author{M.~Stoll}
\affiliation{Institut f{\"u}r Theoretische Physik, Universit{\"a}t
  G{\"o}ttingen, Friedrich-Hund-Platz 1, 37077 G{\"o}ttingen, Germany}

\date{\today}

\begin{abstract}
  The size of the helium trimer is determined by diffracting a beam of
  $^4$He clusters from a 100\,nm grating inclined by 21$^\circ$.  Due to
  the bar thickness the projected slit width is roughly halved to
  27\,nm, increasing the sensitivity to the trimer size. The peak
  intensities measured out to the 8th order are evaluated via a few-body
  scattering theory. The trimer pair distance is found to be $\langle
  r\rangle=1.1\,{+}0.4/{-}0.5\,$nm in agreement with predictions for the
  ground state. No evidence for a significant amount of Efimov trimers
  is found.  Their concentration is estimated to be less than 6\%.
\end{abstract}

 \maketitle

In 1970 Vitali Efimov found a remarkable unexpected property in the
notoriously difficult three-body problem \cite{E_PL33B}. According to
Efimov a weakening of the two-body interaction in a system of three
identical Bosons can lead to the appearance of an infinite number of
bound levels, instead of dissociation as one would expect from classical
mechanics. This effect is related to the divergence of the atomic
scattering length $a$ with decreasing binding energy $E_\text{b}$ between
two of the particles \cite{L_MOLPHYS23}. In nuclear physics, despite extensive
searches, no example for the Efimov effect has been found up to now
\cite{JRFG_RMP76}. At present the most promising candidate is the
{\Heatom} trimer as first predicted by Lim, Duffy, and Damert in 1977
\cite{LDD_PRL38}, although there have been recent attempts to identify Efimov
molecules in ultra-cold collisions of Cs atoms \cite{CVKC_NUCLPA684}.

Because of their very weak binding the existence of the {\Heatom} dimer
and trimer could only recently be established experimentally by a new
technique involving matter-wave diffraction \cite{ST_SCIENCE266}. A beam
of clusters formed in a cryogenic free jet expansion is directed at a
nanostructured $d=100$\,nm period SiN$_x$ transmission grating.  Since
the cluster de Broglie wave length $\lambda$ is inversely proportional
to the cluster number size first order Bragg diffraction peaks for
different sizes are observed at different angles
$\vartheta\approx\lambda/d$, thereby identifying the clusters uniquely.
This technique can also be used to measure the spatial extent of the
clusters. From an analysis of the {\Heatom} dimer diffraction pattern
the slit function of the grating could be determined and from this the
effective slit width for passage of the dimer \cite{GSTHK_PRL83,
  GSTHKS_PRL85}. After accounting for the velocity dependent van der
Waals interaction the effective reduction of the slit width was shown to
be equal to $\frac12\expect{r}$, where the mean bond length was found to
be $\expect{r}=5.2\pm0.4\,$nm \cite{GSTHKS_PRL85}. This extremely large
distance is due to the weak binding energy which was estimated to be
only $|E_\text{b}|=1.1\,{+}0.3/{-}0.2\,$mK \cite{GSTHKS_PRL85}.

\begin{figure}[htbp]
  \centering
  \includegraphics[width=\columnwidth]{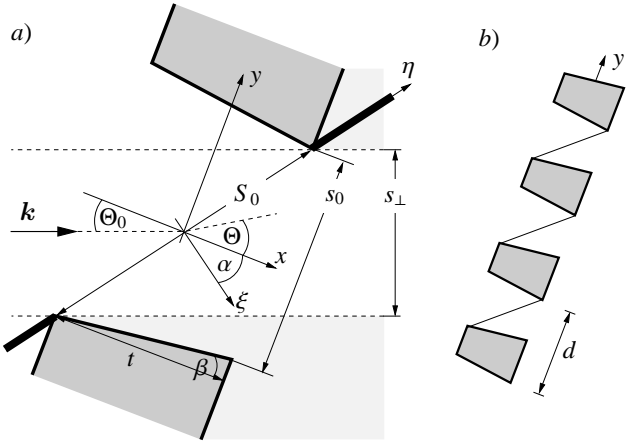}
  \caption{Diffraction geometry at non-normal incidence: \emph{a})
    A single slit of width $\so$ in a plate of thickness $t$ with a
    wedge angle $\beta$ has a projected slit width of $s_\perp$. Both
    the angle of incidence $\Theta_0$ and the angle $\Theta$ are measured
    relative to the plate normal.  The hypothetical thin plate drawn
    along the $\eta$ direction with a slit of width
    $\So=\sqrt{(\so+t\tan\beta)^2+t^2}$ at an angle $\alpha=\arcsin
    (t/\So)$ relative to the thick plate along the $y$ direction casts
    the same geometrical shadow as the thick plate.  \emph{b})
    Transmission grating of period $d$ along the $y$ direction.
    \label{fig:normalnonnormal}}
\end{figure}

For the helium trimer, theory predicts one Efimov state with a similarly
weak binding energy of $|E_\text{e}|=2.3$\,mK in addition to the ground
state with $|E_\text{g}|=126\,$mK with
corresponding pair distances (bond lengths) $\expect{r}=7.97$\,nm and
$0.96$\,nm, respectively \cite{BK_MSSK_BH}. Since for the trimer the slit
width reduction can be shown to be $\frac34\expect{r}$ these two
$s$-states are expected to be distinguishable by their sizes. However,
experiments similar to those used for the dimer did not yield conclusive
results which, ultimately, was attributed to an insufficient resolution.
The present experiment overcomes this limitation by rotating the grating
by an angle $\Theta_0$ around an axis parallel to the slits as seen in
Fig.~\ref{fig:normalnonnormal}.  At $\Theta_0=21^\circ$, due to the
thickness of the bars, the projected slit width is more than halved to
$s_\perp=26.9$\,nm, providing a good compromise between the improvement in
both the ratio $\expect{r}/s_\perp$ and the resolution at the expense of
total transmission. The apparatus used is otherwise similar to the one
described in detail in Ref.~\cite{GSTMSS_PRA61}.  For the trimer
measurements the cryogenic source temperatures $T_0$ and pressures $P_0$
were varied between $(T_0,P_0)=(6.7\,\text{K},1\,\text{bar})$ and
$(40\,\text{K}, 50\,\text{bar})$ to produce optimal trimer mole
fractions of up to 7\% \cite{BST_JCP117}. The collimated beam with a
velocity spread $\Delta v/v\le 2\%$ has a spatial lateral coherence
greater than the exposed 100 grating slits. For both atom and trimer
measurements the mass spectrometer detector was set at the \Heatom$^+$
ion mass. The maximum trimer signal was about $200$ counts/sec.

\begin{figure}[htbp]
  \centering
  \includegraphics[width=\columnwidth]{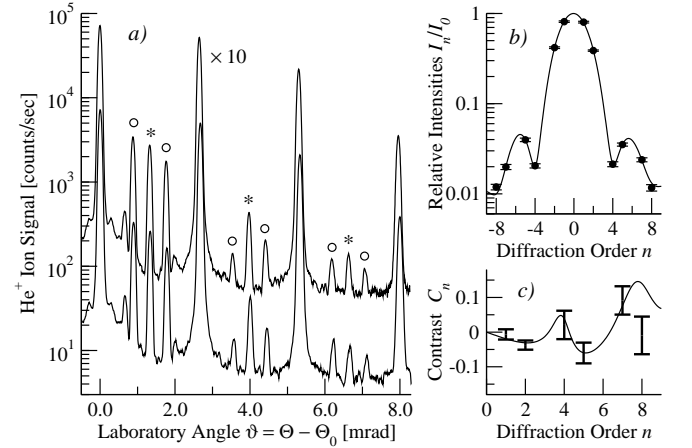}
  \caption{\emph{a}) {\Heatom} diffraction pattern at 
    $\Theta_0=21^\circ$ angle of incidence measured for the source
    conditions $(T_0,P_0)=(16.5\,\text{K}, 7.0\,\text{bar})$
    corresponding to a trimer de Broglie wavelength of
    $\lambda=0.83\,${\AA}. The signal at negative diffraction angles
    has been shifted upwards by a factor 10 and mirrored onto the
    positive side for comparison.  The trimer diffraction peaks are
    marked by circles, dimer peaks by stars.  \emph{b}) Relative
    trimer peak intensities $I_n/I_0$.  \emph{c}) Corresponding
    contrast $C_n=(I_n-I_{-n})/(I_n+I_{-n})$.  The curves in \emph{b})
    and \emph{c}) are best-fit calculations based on
    Eq.~(\ref{eq:diffraction_intensities_approx}).}
  \label{fig:asymmetry}
\end{figure}

Figure \ref{fig:asymmetry}a shows a diffraction pattern out of a
series of altogether 13 taken for various velocities at
$\Theta_0=18^\circ$ and $21^\circ$. The most intense peaks are due to
helium atoms while those marked by circles belong to trimers. Whereas
in the past the diffraction intensities $I_n$ for all orders $n$ had
been found to be perfectly symmetric ($I_n=I_{-n}$), a careful
inspection of the new peak intensities in Fig.~\ref{fig:asymmetry}b
exhibits an up to 10\% deviation from symmetry
\cite{footnote-asymmetry}.  This new feature is clearly demonstrated
by the contrast $C_n=({I_n-I_{-n}})/({I_n+I_{-n}})$ displayed in
Fig.~\ref{fig:asymmetry}c. By modifying the diffraction theory of
Ref.~\cite{HK_PRA57_61} to account for the asymmetry the new
measurements could be evaluated to obtain the bond length of the
helium trimer $\expect{r}=1.1\, {+}0.4/{-}0.5\,$nm.  Assuming the
theoretical values of $\expect{r}$ the maximum concentration of Efimov
trimers in the beam is estimated to be less than 6\%.  Although their
small concentration may possibly be explained by collisional depletion
in the expansion, the negative result obviously also raises new doubts
about the existence of an Efimov state in {\Hetrimer}.

From atom beam transmission experiments \cite{GSTMSS_PRA61} the
grating bars are found to have a thickness $t=118.3\pm0.5$ nm and
their inner faces have a wedge angle $\beta=6.7\pm 0.5^\circ$ with the
direction perpendicular to the grating
(Fig.~\ref{fig:normalnonnormal}).  Since the angle of inclination
(angle of incidence) $\Theta_0$ exceeds the wedge angle $\beta$ the
upper bar faces (Fig.~\ref{fig:normalnonnormal}) are shadowed by the
front edges of the bars.  Obviously in this geometry the opening
($\so$ in Fig.~\ref{fig:normalnonnormal}) used in previous
calculations of the scattering amplitude for normal incidence
\cite{GSTHK_PRL83, GSTHKS_PRL85} is no longer appropriate.  Instead
the slit is modeled by a diagonal opening of width $\So$ in a thin
plate along the $\eta$ axis (Fig.~\ref{fig:normalnonnormal}) which
casts the same geometrical shadow as the original slit
\cite{footnote-formalderiv}.  Complications from scattering from the
upper bar faces are not expected since the cluster de Broglie wave
length $\lambda\approx1\,${\AA} is much smaller than the slit width
such that the diffraction is concentrated in a small range of angles
of the order of $\vartheta=\Theta-\Theta_0\simeq
\lambda/s_\perp\approx 2^\circ$, much smaller than
$\Theta_0-\beta\approx 14^\circ$.  Modeling the incident beam by a
plane wave of wave vector $\bm{k}$ with $k=|\bm{k}|=2\pi/\lambda$ and
imposing Kirchhoff boundary conditions along the slit $\So$ leads to
the following expression for the scattering amplitude of the diagonal
slit \cite{bornwolf,GSTHK_PRL83}
\begin{equation}
  \label{eq:scattering_amplitude_slit}
  f_\text{slit}(\Theta)=\frac{\cos(\Theta_0+\alpha)}{\sqrt{\lambda}}
  \int_{-\So/2}^{\So/2}d \eta\ e^{-i K(\Theta) \eta}\ \tau(\eta),
\end{equation}
where the bar thickness enters through $\sin\alpha=t/\So$ and
\begin{equation}
  \label{eq:Deltaketa}
  K(\Theta)=k \left[\sin(\Theta+\alpha)-\sin(\Theta_0+\alpha)\right]
\end{equation}
is the wave vector transfer along the slit direction ($\eta$ axis).
The transmission function $\tau(\eta)$ in
Eq.~(\ref{eq:scattering_amplitude_slit}) accounts for the size of the
cluster \cite{GSTHKS_PRL85} as well as the weak van der Waals surface
interaction of the form $-C_3/l^3$ between the atoms and the bar
material \cite{GSTHK_PRL83}, where $l$ is the distance from the
surface.

The inclined transmission grating consists of many slits aligned along
the $y$ axis with period $d$ (Fig.~\ref{fig:normalnonnormal}b).  The
periodicity then gives rise to sharp principal diffraction maxima
located at the Bragg angles $\Theta_n$ satisfying
\begin{align}
  \label{eq:Deltaky}
  k\left[\sin\Theta_n-\sin\Theta_0\right]=2\pi n/d
\end{align}
for $n=0,\pm1,\pm2,$ etc \cite{footnote-braggapprox}.  Solving
Eq.~(\ref{eq:Deltaky}) for $\Theta_n$ and inserting into
Eq.~(\ref{eq:Deltaketa}) yields $K(\Theta_n)$ at which the scattering
amplitude determining the intensity of the $n$-th diffraction order is
to be evaluated.  By expanding through second order in $n$,
$K(\Theta_n)$ can be expressed as
\begin{align}
  \nonumber
  \frac{K(\Theta_n)}{\cos(\Theta_0+\alpha)}\!\approx\!
  \frac{2\pi n}{d\cos\Theta_0}
  +\lambda\ \frac{\tan\Theta_0\!-\tan(\Theta_0+\alpha)}{4\pi}
  \left(\frac{2\pi n}{d\cos\Theta_0}\right)^2\!\!.
\end{align}
Thus, although the scattering amplitude itself is even under the
change of the sign of $K(\Theta)$, it is probed at a wave vector
transfer for which $K(\Theta_n)\neq -K(\Theta_{-n})$. The origin of the
experimentally observed asymmetry $I_n\neq I_{-n}$ of the diffraction
pattern lies, therefore, in the non-alignment of the slits $\So$
($\eta$ axis) and the direction of periodicity ($y$ axis).  The
asymmetry decreases with $\lambda$ because for a smaller de Broglie
wave length less clusters are diffracted into the shaded region of
the slits (Fig.~\ref{fig:normalnonnormal}).  Clearly, for $\alpha=0$
(thin grating) the symmetric case is recovered. Supplementary
calculations indicate that the van der Waals surface interaction has
only a minor effect on the asymmetry.

An analytical expression for the relative diffraction intensities
$I_n/I_0$ is obtained by introducing the functions $\Phi^+(K)$ and
$\Phi^-(K)$ \cite{GSTHK_PRL83}
\begin{align}
  \Phi^\pm(K)=
  \int_0^{\So/2}d\eta\ e^{\pm iK\eta}\
  \frac{  \frac{\partial}{\partial \eta}\tau[\pm(\So/2-\eta)]}{\tau(0)}
\end{align}
which allows the scattering amplitude to be expressed exactly as
\begin{align}
  \nonumber
  f_\text{slit}(\Theta)&=\frac{\cos(\Theta+\alpha)}{\sqrt{\lambda}}
  \ \tau(0)\\
  \label{eq:scattering_amplitude_slit-kum}
  &\times 
  \frac{e^{iK(\Theta)\So/2}\Phi^-(K(\Theta))
    -e^{-iK(\Theta)\So/2}\Phi^+(K(\Theta))}{iK(\Theta)}.
\end{align}
To conveniently combine the functions $\Phi^\pm(K)$ with the
exponentials in Eq.~(\ref{eq:scattering_amplitude_slit-kum}) their
logarithms are expanded in a power series:
$\ln\Phi^\pm(K)=\sum_{j=1}^\infty {(\pm iK)^j}R^\pm_j/j!$, which
uniquely defines the complex numbers $R^\pm_j$ known as the cumulants.
For example, the first cumulants are given by
$R^\pm_1=\pm\int_0^{\pm\So/2} d\eta\ \left[1-\tau(\eta)\right]$ and
account for the different transmission in the two halves of the slit.
For the diffraction orders $|n|\lesssim 8$ encountered experimentally
it is sufficient to retain only the first two terms of this expansion.
Inserting them into Eq.~(\ref{eq:scattering_amplitude_slit-kum}) the
$n$-th order diffraction intensity becomes, to good approximation,
\begin{align}
  \label{eq:diffraction_intensities_approx}
  \frac{I_n}{I_0}= \frac{e^{-K(\Theta_n)^2\Sigma^2}e^{-K(\Theta_n)\Gamma}}
  { \Big( \frac{K(\Theta_n)\sqrt{S_\text{eff}^2+\Delta^2}}{2} \Big)^2 }
  \left[
    \sin^2\left(\frac{K(\Theta_n) S_\text{eff}}2\right)
    +\sinh^2\left(\frac{K(\Theta_n) \Delta}2\right) \right].
\end{align}
Here, the effective slit width $S_\text{eff}=\So-\Re(R^+_1+R^-_1)$
accounts for the reduction of the geometrical slit width $\So$ due to
the surface interaction as well as the finite cluster size.  The
exponential involving $\Sigma=\sqrt{\Re(R^+_2+R^-_2)/2}\approx5\,$nm
includes the Debye-Waller attenuation due to irregular variations of
the slit width across the grating and also accounts for cluster
breakup \cite{GSTHK_PRL83,GSTHKS_PRL85}.  The surface interaction
removes the intensity zeros through the term involving
$\Delta=\Im(R^+_1+R^-_1)\approx 10\,$nm and contributes weakly to the
asymmetry through $\Gamma=\Im(R^+_1-R^-_1)\approx 1.5\,$nm.

\begin{figure}[htbp]
  \centering
  \includegraphics[width=\columnwidth]{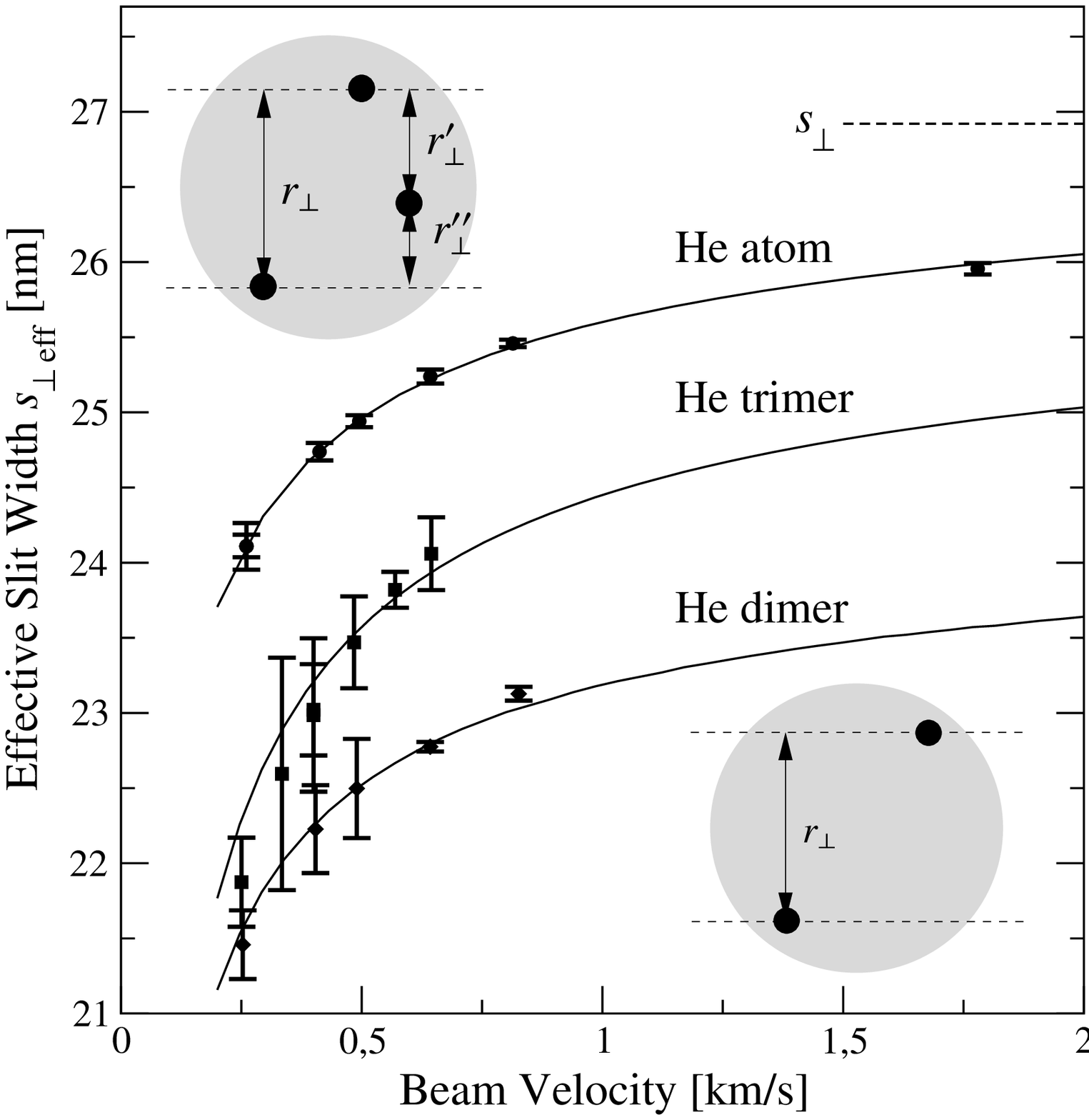}
  \caption{Projected effective slit widths
    $s_{\perp\text{eff}}=S_\text{eff}\cos(\Theta_0+\alpha)$ measured
    at different beam velocities at an angle of incidence
    $\Theta_0=21^\circ$ for {\Heatom}, {\Hedimer} and {\Hetrimer}. The
    curves represent best fits of
    Eq.~(\ref{eq:effective_slit_perp_width_trimer}) for {\Hetrimer}
    and analogous expressions for {\Heatom} and {\Hedimer}. Their high
    velocity limits are given by $s_\perp$ for {\Heatom}, by
    $s_\perp-\frac12\expect r$ for {\Hedimer}, and by
    $s_\perp-\frac34\expect r$ for {\Hetrimer}. The insets illustrate
    the ``widths'' traced out by the clusters along their flight paths
    (see text).}
  \label{fig:effective_slit_width_data}
\end{figure}

Experimental values for $S_\text{eff}$ were obtained from fits of the
intensity formula Eq.~(\ref{eq:diffraction_intensities_approx}) to
trimer diffraction patterns (Fig.~\ref{fig:asymmetry}b) measured for
$T_0=6.7$ -- $40\,$K. In Fig.~\ref{fig:effective_slit_width_data} the
projected effective slit widths
$s_{\perp\text{eff}}=S_\text{eff}\cos(\Theta_0+\alpha)$ for {\Heatom},
{\Hedimer} and {\Hetrimer} at $\Theta_0=21^\circ$ are plotted as
functions of the beam velocity.  The atom data were used to determine,
along the lines of Ref.~\cite{GSTHK_PRL83}, the projected slit width
$s_\perp=26.92\pm0.02$\,nm and the van der Waals interaction
coefficient was taken as $C_3=0.113\pm0.02$ meV nm$^3$
\cite{GSTHK_PRL83}. As seen from
Fig.~\ref{fig:effective_slit_width_data} the trimer size effect at a
velocity of $0.64\,$km/s is of the order of only $1.2\,$nm, clearly
smaller than the $2.5\,$nm for the dimer.  Moreover, the dimer curve
runs almost parallel to the atom curve suggesting that, due to the
extent of the dimer wave function, on average only one of its atoms is
interacting with the surface. In contrast, the steeper slope of the
trimer curve indicates the contribution of more than one atom, also
confirming the relative compactness of this cluster.

The smallness of the trimer binding energy as compared to its kinetic
energy allows an extension of the quantum mechanical few-body
scattering approach of Ref.~\cite{HK_PRA57_61}, which is based on the
AGS equations \cite{AGS} and originally designed to describe dimer
diffraction, to treat also the trimer case. It turns out that, as a
generalization of the dimer result, the size effect is caused by the
width of the trimer perpendicular to its incident direction.
For the dimer \cite{GSTHKS_PRL85} this width can be expressed
by the expectation value $\expect{|r_\perp|}=\expect{r}/2$ where $r$
denotes the dimer bond length and $|r_\perp|$ is its perpendicular
projection (lower inset in Fig.~\ref{fig:effective_slit_width_data}).
The analogous expression for the trimer, which is more intricate due
to the third atom, is given by $(|r_\perp|+|r'_\perp|+|r''_\perp|)/2$
where the three distances are defined in the upper inset of
Fig.~\ref{fig:effective_slit_width_data}. For the homonuclear
{\Hetrimer} the expectation value of this quantity reduces to
$3\expect{|r_\perp|}/2$. Moreover, since the pair interactions are
dominated by the shallow $s$-wave dimer state, the homogeneous Faddeev
equations \cite{sitenko} can be used to express the width in terms of
the trimer bond length as $3\expect{r}/4$. The complete expression for
$s_{\perp\text{eff}}$, which includes the surface interaction, is then
found to be
\begin{align}
  \nonumber
  s_{\perp\text{eff}}&= s_\perp-\frac34\expect{r}-
  \zeta\Re\\
  \nonumber
  &\times
  \left\{
    \int_0^{\So/2} \!\!\!\!\!d \eta
    \left[1-\tau_\text{at}(\eta)
      \tau_\text{at}\left(\eta-\textstyle{\frac12
          \frac{\expect{r}}{\zeta}}\right)
      \tau_\text{at}\left(\eta-\textstyle{\frac58
          \frac{\expect{r}}{\zeta}}\right)
    \right]
  \right.\\
  &\ \ \ +\!\!
  \left.
    \int_{-\So/2}^0\!\!\!\!\!d \eta
    \left[1-\tau_\text{at}(\eta)
      \tau_\text{at}\left(\eta+\textstyle{\frac12
          \frac{\expect{r}}{\zeta}}\right)
      \tau_\text{at}\left(\eta+\textstyle{\frac58
          \frac{\expect{r}}{\zeta}}\right)
    \right]
  \right\}.
  \label{eq:effective_slit_perp_width_trimer}
\end{align}
were $\zeta=\cos(\Theta_0+\alpha)$ was used. The term in curly braces
in Eq.~(\ref{eq:effective_slit_perp_width_trimer}) accounts for the
surface interaction via the atom transmission functions
$\tau_\text{at}(\eta)$ \cite{GSTHK_PRL83}. As seen in
Fig.~\ref{fig:effective_slit_width_data} this term varies between
2--4\,nm in the experimental range of 0.25 -- 0.64\,km/s.

Using Eq.~(\ref{eq:effective_slit_perp_width_trimer}) the best fit
curve for the trimer based on seven diffraction patterns taken at
$\Theta_0=21^\circ$ was obtained for the bond length
$\expect{r}=1.0\,{+}0.5/{-}0.7$\,nm. A second series of six
diffraction patterns taken at $\Theta_0=18^\circ$ yielded
$\expect{r}=1.2\,{+}0.5/{-}0.8$\,nm which confirms, within the error
bars, the reproducibility of the result.  The average of both results,
$\expect{r}=1.1\,{+}0.4/{-}0.5$\,nm, agrees well with the theoretical
prediction of $0.96$\,nm \cite{BK_MSSK_BH} for the {\Hetrimer} ground
state.  Moreover, it rules out a significant concentration of the
Efimov state in the beam. With the theoretical values for $\expect{r}$
a simulation of the diffraction pattern for various concentrations
indicates that the upper experimental limit is consistent with less
than 6\% Efimov trimers, reducing substantially the previous value of
15\% \cite{KKRTV_PRL93}.

Dedicated calculations for the formation of excited state clusters
during the beam expansion are not available. Applying, for an
estimate, the equilibrium model of Ref.~\cite{BST_JCP117} the ratio of
Efimov to ground state trimers is expected to be approximately
proportional to $\exp(-|E_\text{g}-E_\text{e}|/k_\text{B}T_\infty)$.
At an asymptotic temperature in the fully expanded beam of
$T_\infty=1\text{--}5\,$mK in the present experiment, this is indeed a
very small number $(< 10^{-11})$.  The validity of this estimate,
however, depends on the temperature at which the internal states
equilibrate, which may be much larger than $T_\infty$. While the
collisional de-excitation of an Efimov trimer into the ground state is
expected to be small \cite{HK_PRL84} a realistic calculation would
need to take into account the inelastic cross-section between the
Efimov trimer and the co-expanding atoms.  Indirect evidence for the
likely robustness of the Efimov cluster comes only from the large mole
fraction of the even more weakly bound {\Hedimer} \cite{GSTHKS_PRL85}.

There is, of course, also the possibility that the {\Hetrimer} Efimov
state does, in fact, not exist despite the over 40 publications which
have appeared since 1977. Since all calculations have been carried out
for adiabatic two-body potentials, which have been tested both
experimentally \cite{GSTHKS_PRL85} and by numerical methods
\cite{A_JCP120}, it is still conceivable that the presence of the
Efimov state is affected by the sum of so far neglected small
corrections to the potentials, such as a three-body contribution to
the interaction \cite{RA_JCP102}, retardation or non-adiabatic effects
\cite{G_MOLPHYS99}. For example, Gdanitz \cite{G_MOLPHYS99} showed
that the latter can modify the scattering length by about 5--10\%.
However, a solution of the Faddeev equations based on a separable
potential, which was adjusted to reproduce exactly the scattering
length and the effective range of the He-He interaction, reveals that
such a modification alone cannot render the Efimov state unbound.

In future experiments a promising approach to detect Efimov
{\Hetrimer} could involve sampling the sizes of clusters effusing from
a Knudsen cell, thereby ruling out collisional de-excitation. Then the
Efimov and ground state molecules would have small but nearly equal
concentrations. To compensate for the loss in signal the trimer mole
fraction could be increased by going to a much higher $P_0$ while
reducing the orifice diameter. A new, much more sensitive detector
currently under development may make such experiments possible.

\begin{acknowledgments}
  We are indebted to T.~Savas for providing the transmission grating
  and thank T.~K{\"o}hler for stimulating discussions.
\end{acknowledgments}


\begin{thebibliography}{23}
\bibitem{footnote-asymmetry} {As seen in Fig.~\ref{fig:asymmetry} the
    diffraction angles $\vartheta_n=\Theta_n-\Theta_0$ are also asymmetric
    ($\vartheta_n\neq-\vartheta_{-n}$) in agreement with the Bragg law
    Eq.~(\ref{eq:Deltaky}). However, this effect is unrelated to the
    intensity asymmetry.  }
\bibitem{footnote-braggapprox}
  {For $n2\pi/(kd)\ll 1$ one has $\vartheta_n=\Theta_n-\Theta_0\approx
    n2\pi/(kd\cos\Theta_0)$, as used in Ref.~\cite{GSTMSS_PRA61}.}
\end{thebibliography}
\end{document}